\documentclass[a4paper]{jpconf}
\usepackage{graphicx}
\begin{document}

\title{Thermoelectric transport in two-dimensional topological insulator state based on HgTe quantum well}

\author{G. M. Gusev,$^1$  O. E. Raichev,$^2$  E. B. Olshanetsky,$^3$  A. D. Levin,$^1$ Z. D. Kvon,$^{3,4}$ N. N. Mikhailov,$^3$ and S. A. Dvoretsky,$^{3}$}

\address{$^1$Instituto de F\'{\i}sica da Universidade de S\~ao
Paulo, 135960-170, S\~ao Paulo, SP, Brazil}
\address{$^2$Institute of Semiconductor Physics, NAS of
Ukraine, Prospekt Nauki 41, 03028 Kyiv, Ukraine}
\address{$^3$Institute of Semiconductor Physics, Novosibirsk
630090, Russia}
\address{$^4$Novosibirsk State University,
Novosibirsk, 630090, Russia}

\ead{gusev@if.usp.br}

\begin{abstract}

The thermoelectric response of HgTe quantum wells in the state of
two-dimensional topological insulator (2D TI) has been studied
experimentally. Ambipolar thermopower, typical for an
electron-hole system, has been observed across the charge
neutrality point, where the carrier type changes from electrons to
holes according to the resistance measurements. The hole-type
thermopower is much stronger than the electron-type one. The
thermopower linearly increases with temperature. We present a
theoretical model which accounts for both the edge and bulk
contributions to the electrical conductivity and thermoelectric
effect in a 2D TI, including the effects of edge to bulk leakage.
The model, contrary to previous theoretical studies, demonstrates
that the 2D TI is not expected to show anomalies of thermopower
near the band conductivity threshold, which is consistent with our
experimental results. Based on the experimental data and
theoretical analysis, we conclude that the observed thermopower is
mostly of the bulk origin, while the resistance is determined by
both the edge and bulk transport.

\end{abstract}

\section{Introduction}

The HgTe-based quantum well (QW) is a semiconductor system, where
the two-dimensional (2D) conduction and valence subbands, divided
by a narrow variable gap, can be created. The switch between the
electron and the hole types of transport is achieved by the gate
control. The gap energy is controlled by the well width $d$ and
goes to zero at $d \simeq 6.3$ nm, when the electron energy
spectrum resembles a Dirac cone, like in graphene. Wider wells
have an inverted energy band order, known to be in the state of 2D
topological insulator (TI) [1] characterized by a pair of
counterpropagating gapless edge modes. The edge states have a
helical spin structure and are supposed to be robust to
backscattering. The edge state transport in such HgTe QWs has been
confirmed experimentally for ballistic transport in mesoscopic
samples [2-4] and for diffusive transport in macroscopic samples
[4,5]. Similar results have been obtained in experiments on
Si-doped InAs/GaSb quantum wells [6-9] which are also believed to
be 2D TI. Application of novel experimental methods for the study
of the transport properties of 2D TI is of particular interest.

The thermoelectric measurements can give complementary information
about electron transport in metals and semiconductors and are used
as a powerful tool for probing the sign of the charge carriers and
the transport mechanisms. The diffusive thermopower is often
described using the Mott relation, as the logarithmic derivative
of the energy-dependent electrical conductivity. Apart from
necessitating the validity of the Boltzmann equation and the
degeneracy of the electron gas, the Mott relation has also other
limitations discussed in the literature. In particular, a strong
energy dependence of the relaxation time, when this time changes
considerably within the $kT$ interval around the Fermi level,
causes a failure of the simple Mott relation. Indeed, the
deviation from the Mott relation has been observed in thermopower
experiments in graphene near the charge neutrality point (CNP) and
at high temperatures [10,11]. Application of the original Mott
relation [12], which is more general because it is not based on
the approximate Sommerfeld expansion, allows one to overcome these
limitations and describe the thermopower in a wide range of
temperature and carrier density [13]. However, in the case of
strongly inelastic scattering by optical phonons, considerable
deviations even from the Mott relation in its general form are
expected, as recently demonstrated in high mobility graphene
samples [14,15].

Similar to graphene, the system based on HgTe quantum wells
reveals ambipolar Hall effect accompanied by the resistance peak
near the CNP. However, in contrast to the gapless graphene, in 2D
TI the transition between the electron and hole types of
conduction as the gate voltage is swept through the CNP, occurs
when the Fermi level stays in the insulating gap and transport is
determined by the edge states. The position of the Fermi level in
the gap is stabilized by the bulk states which are present in the
gap because of a random spatial inhomogeneity (disorder). These
states are often considered as localized ones. However, electron
transitions between the edge and the bulk states are possible and
may influence transport properties in 2D topological insulators
[4,5,16]. The edge to bulk mixing is believed to cause a strong
enhancement of the thermopower in 2D TI. In particular, it is
suggested [17,18] that when the Fermi level approaches the bulk
band edge, the scattering rate of electrons in the edge states
increases rapidly and significantly, which is expected to cause an
anomalous growth of the amplitude of Seebeck signal and a change
of its sign. This offers a new opportunity to improve the
thermoelectric parameter, such as the figure of merit $zT$, which
is defined as $zT=G S^{2}T/(K_{e}+K_{ph})$, where $G$ is the
electron conductance, $S$ is the Seebek coefficient, $K_{e}$ and
$K_{ph}$ are the thermal conductances of electrons and phonons
consequently. The interplay between the edge and the bulk
conductances leads to a strong dependence of the parameter $zT$ on
the sample geometry and size. It has been predicted that the value
of the figure of merit can be improved by more than $\sim 1$ for a
certain geometry at room temperature [17,18]. Despite the interest
to the thermoelectric properties in 2D topological insulators, the
experimental studies have almost all been focused on the
measurements of the electrical resistance.

In the present paper we report an experimental study of the
thermopower in band-inverted HgTe-based quantum wells. At the CNP
where the resistance reaches its maximum, the thermopower changes
its sign, showing the ambipolar behavior. The nonlocal resistance
in our samples is comparable with the local one. This observation
clearly proves the presence of the edge state transport, which
dominates within the bulk gap. Importantly, we do not observe any
of the anomalies of the Seebeck effect predicted in Refs. [17,18],
in particular, the sign of the effect changes like in a normal
electron-hole system. This apparently suggests that the effect of
the edge to bulk scattering on the transport is not as significant
as it was expected. To verify this statement, we have carried out
a calculation of conductivity and thermopower in the 2D TI, taking
into account both the particle and energy balance in the coupled
system of edge and bulk states. In brief, we demonstrated that the
transport properties are determined not by the edge to bulk
scattering rate alone, but by the spin current flowing between the
edge and the bulk. If the spin relaxation in the bulk is slow, a
bottleneck effect takes place, when the spin current is limited by
the bulk conductivity rather than by the edge to bulk scattering
rate. Therefore, in the region where the bulk conductivity is
smaller than the conductance quantum $e^2/h$ the scattering
between the edge and the bulk is expected to be insignificant,
while in the regions of larger bulk conductivity the edge-state
contribution to transport is no longer important. Further, from a
qualitative analysis of our experimental data supported by the
theoretical considerations, we conclude that the observed
thermopower is mostly of the bulk transport origin.

The paper is organized as follows. Section II contains the
description of measurements and experimental results. Section III
is devoted to the theoretical model and to discussion of the
results based on this model. The concluding remarks are given in
the last section.

\begin{figure}[ht!]
\includegraphics[width=15cm,clip=]{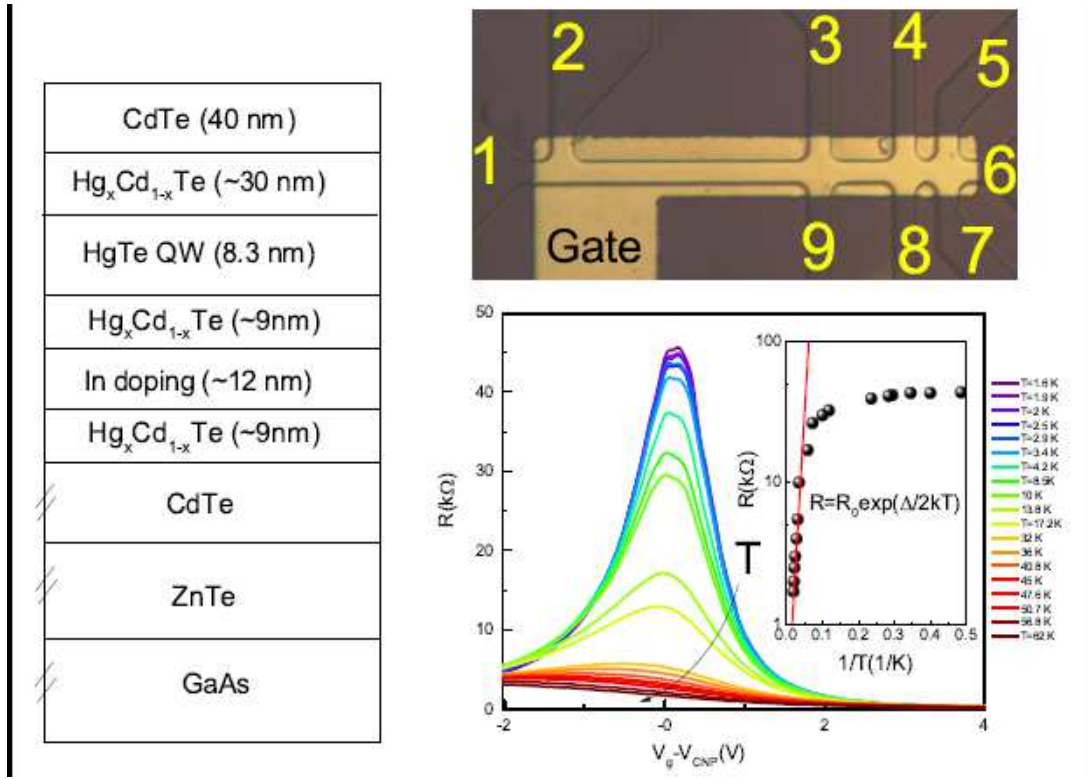}
\caption{\label{fig.1}(Color online) Sample geometry and resistance of shortest segment ($R_{I=1,6; V=4,5}$)
 as a function of the gate
voltage for different temperatures (Sample A). Left-schematic structure of the sample.}
\end{figure}

\section{Experiment}

To probe the carrier transport, we measured thermoelectric voltage
$V$ and thermopower $S$ together with the resistance. The quantum
well structures
Cd$_{0.65}$Hg$_{0.35}$Te/HgTe/Cd$_{0.65}$Hg$_{0.35}$Te with [013]
surface orientations and widths $d=8-8.3$ nm were prepared by
molecular beam epitaxy (figure 1, left panel). The sample is a
long Hall bar consisting of three 3.2 $\mu$m wide consecutive
segments of different length (3, 9, and 35 $\mu$m) and seven
voltage probes, covered by the TiAu gate ( see figure 1, top
panel). The measurements were performed in a variable temperature
insert cryostat in the temperature range $1.4-10$ K using the
standard four point scheme. A detailed description of the sample
structure has been given in Ref. [5]. The electrically powered
heater placed symmetrically near the contact 1 (see Fig. 2, top
panel) creates temperature gradient in the system, while the other
end is indium soldered to a small copper slab that serves as a
thermal ground. The copper slab is, in turn, connected to the
copper rod of the sample holder. One calibrated  thermo sensor is
attached at the end of the sample near the heater while the other
is attached to the
 heat sink. The thermo sensors were used to measure the $\Delta T$
along the sample. The voltages induced by this gradient were
measured by a lock-in detector at the frequency of $2f_{0}=0.8-2$
Hz across various voltage probes. The thermal conductance of the
sample is overwhelmingly dominated by phonon transport in the GaAs
substrate [19,20]; diffusive heat transport by the two-dimensional
gas is negligible in comparison. The thermal conductivity $\kappa$
of a pure dielectric crystal is usually determined by the boundary
scattering at low temperatures and depends on the temperature as
$\kappa \sim T^{3}$. We performed the measurements of the thermal
conductivity in our samples and obtained the  value $300-200 W
m^{-1}K{-1}$ at T=4.2K for different substrates, which agrees with
the previous measurements [19,20]. We did not directly measure the
temperature difference between the voltage probes, since the
distance is very small. We estimate this difference between probes
3 and 2 as $\sim20 mK$ K for the heater power used in our
experiment. For a given temperature difference between the sample
extremities the temperature profile along the sample could be
nonlinear, especially close to the ends of the substrates. The
situation is somewhat similar to the electrical measurements, when
the electric potential profile is inhomogeneous near the metallic
contacts. However, we expect that in the center of the sample and
along the short distances the profile is linear. We have also
checked that the temperature difference varies linearly with
heater power. Four different devices have been studied. Below we
show the results obtained in two representative samples ( A and
B).

\begin{figure}[ht!]
\includegraphics[width=15cm,clip=]{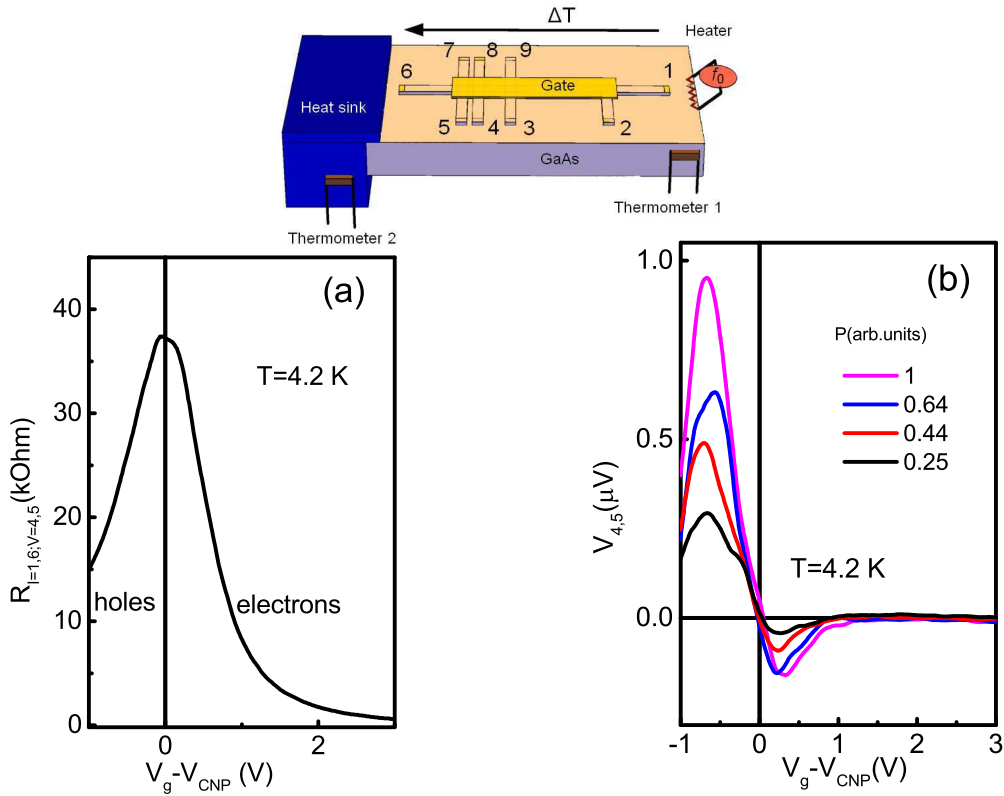}
\caption{\label{fig.2}(Color online) Resistance (a) and
thermovoltage  (b) for different heater powers (indicated) as a
function of gate voltage , $T=4.2$ K (sample A).}
\end{figure}
The variation of the resistance with the gate voltage and lattice
(bath) temperature is shown in Figure 1. The resistance of the
shortest segment reveals a broad peak whose amplitude is larger
than the value $h/2e^2$ expected in the ballistic case. We  see
that the resistance decreases sharply for temperatures above 15 K
while saturating below 10 K. We  find  that  the  profile  of  the
resistance temperature dependencies  above $T > 15 K$  fits  very
well  the activation law $R~exp(\Delta/2 kT)$, where $\Delta$ is
the activation gap. Insert in Figure 1 shows  the peak maximum
resistance versus temperature. The  thermally activated behavior
of resistance  above  15  K corresponds  to a gap  of 10 meV
between the  conduction and  valence  bands in the HgTe well. The
mobility  gap  can  be  smaller than the energy gap due  to
disorder.
\begin{figure}[ht!]
\includegraphics[width=15cm,clip=]{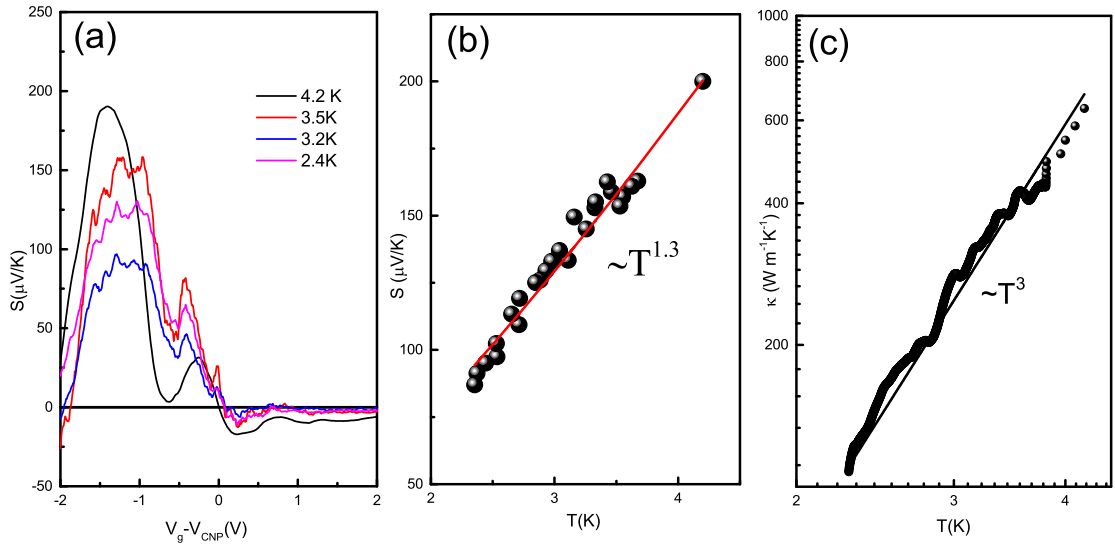}
\caption{\label{fig.3}(Color online) (a) Thermopower for different temperatures. (b) Temperature dependence of thermopower  at $V_{g}-V_{CNP}=-1.2$ V.
(c) Lattice thermoconductivity of the GaAs substrate as a function of the temperature (Sample B).}
\end{figure}
Figure 2 shows the resistance and thermovoltage as a function of
the gate voltage measured between  probes 4 and 5 at $T=4.2$ K.
The thermovoltage increases nearly linearly with heater power,
which proves that we measure the longitudinal (Seebeck)
thermoelectric effect. The signal shows a behavior similar to
other electron-hole systems such as graphene [10,11]. It changes
sign at the charge neutrality point (CNP) and decreases with the
carrier density increasing. The voltage interval between the
electron-like and hole-like regimes ($\Delta V_g \sim 1$ V) is
almost two times smaller than the half-width of the resistance
peak.
 The Figure 3a displays the traces of the thermopower versus $V_{g}$ for different
temperatures. Figure 3b shows the temperature dependence of the
thermopower measured across a longer bridge at a selected gate
voltage $V_{g}-V_{CNP}=-1.2$ V (hole side) where the thermopower
approaches its maximum.  It is found that the signal grows almost
linearly with temperature: $S\sim T^{1.3\pm0.1}$ ( figure 3b) in
the temperature interval $2.2 < T < 4.2K$. This temperature
interval was selected, because resistance becomes temperature
dependent above $T > 10 K $( see figure 1), and metallic
approximation for thermopower is no longer valid. It is worth
noting that prior to the thermoelectric measurements the thermal
conductance of the sample has been determined. The
thermoconductance is dominated by the phonon transport in the
substrate; the contribution from the diffusive heat transfer by
the electrons is negligibly small. The thermal conductivity of the
GaAs substrate is usually determined by the boundary scattering at
low temperatures [19, 20]. The thermoconductivity is given by:
\begin{equation}
\kappa=\frac{2\pi^{2}}{15}\frac{k\Lambda}{v_{ph}^{2}}\left(\frac{kT}{h}\right)^{3},
\end{equation}
where $\Lambda$ is the phonon mean-free path and $v_{ph}=3300 m/s$
the appropriate mean acoustic phonon velocity. Figure 3c shows the
thermal conductivity of the GaAs substrate as a function of the
lattice temperature. The thermal conductivity follows $\kappa\sim
T^{3}$ law, in accordance with eq.1. The value of $\kappa$ is
$\sim 600 Wm^{-1}K^{-1}$ at T=4.2K, which agrees well with the
previously measured thermoconductivity in pure GaAs substrates
[19,20]. Once the thermal conductance is found, the temperature
gradient can be used to convert the measured thermoelectric
voltages into the thermopower.
\begin{figure}[ht!]
\includegraphics[width=15cm,clip=]{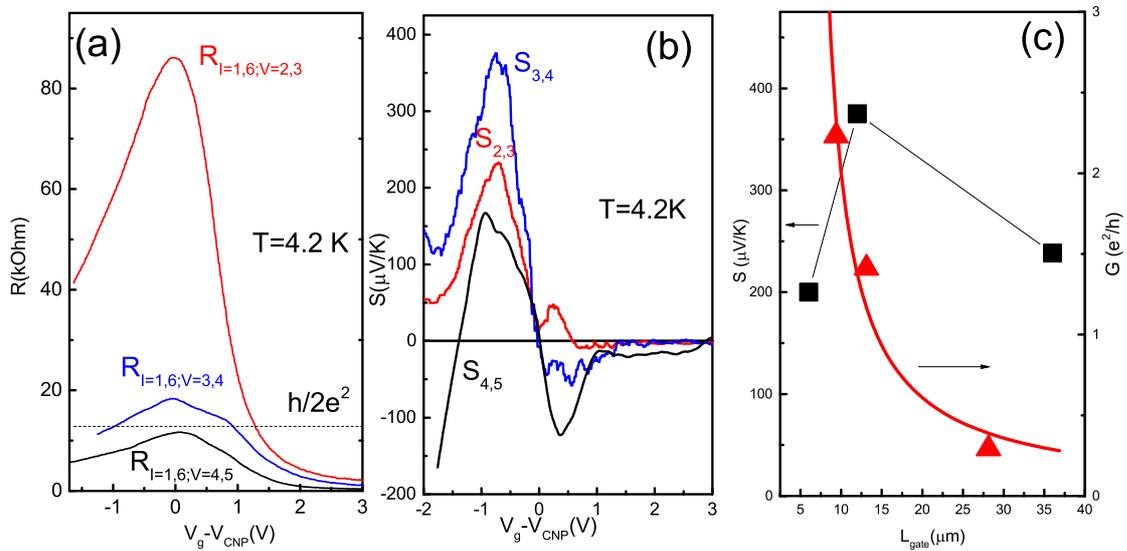}
\caption{\label{fig.4}(Color online) (a) Resistance $R$ as a
function of gate voltage measured between various voltage probes,
$T=4.2$ K, $I=10^{-9}$ A. (b) Thermopower as a function of the
gate voltage measured between various voltage probe, $T=4.2$ K.
(c) Resistance at the CNP and thermopower at $V_{g}-V_{CNP}=-0.7V$
as a function of the distance between the voltage probes $L$
(Sample B). }
\end{figure}
Figure 4 shows resistance as a function of the gate voltage measured between different probes at $T=4.2$ K.
 It is worth noting that the edge current flows along the gated
sample edge whose length $L_{gate}$ is longer than the distance
between the probes $L$ (bulk current path) and corresponds to 5-6
$\mu$m. For longer distances between the probes we see higher
resistances. The large resistance can appear because of multiple
transitions of electrons between counterpropagating helical edge
states caused by either direct backscattering or electron transfer
mediated by the bulk states in the puddles [21] emerging due to
spatial potential fluctuations near the edge. The observation of a
nonlocal resistance constitutes the main proof of the presence of
the edge state transport in a 2D TI. A systematic study of the
local and nonlocal transport in 2D TI has been preformed in the
previous works in the ballistic [3,4] and in the diffusive regimes
[16]. The dependence of the resistance peak on the length
$L_{gate}$ is shown in figure 4c. It is found to be very close to
the $1/L_{gate}$ dependence. The thermopower signal has a
nonmonotonic dependence on the distance $L_{gate}$ (figure 4c) in
contrast to the resistance. Below we consider the theory, which
accounts for both the edge and bulk contributions to the
electrical conductance and thermoelectric effect in 2D TI,
including the effects of edge to bulk leakage.

Finally a few words need to be said about the thermoelectric
efficiency in the 2D topological insulator regime. As mentioned in
the introduction, in conventional semiconductors the factor $zT$
is size independent because the geometrical factor is canceled
between the conductance and the thermoconductance. In 2D
topological insulator regime factor $zT$ can be optimized by
choosing appropriate geometries [18]. A large enhancement of the
power factor is predicted for the gapless edge states near the
charge neutrality point. Figure 5a shows the conductance measured
near the CNP as a function of the gate voltage. It is expected
that when the edge-state contribution to the transport is
important, the nonlocal resistance should be observed [3,4].
\begin{figure}[ht!]
\includegraphics[width=12cm,clip=]{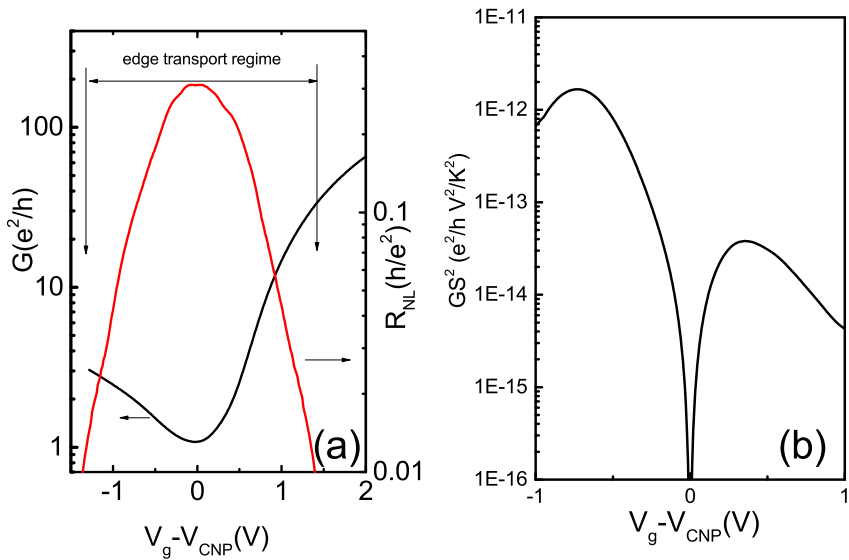}
\caption{\label{fig.5}(Color online) (a) Conductance $G=1/R_{I=1,6;V=4,5}$ and nonlocal resistance $R_{NL}=R_{I=4,8;V=5,7}$ as a function of the gate voltage.
(b) Coefficient $GS^{2}$ as a function of the gate voltage near CNP, T=4.2K (Sample B).}
\end{figure}

For comparison figure 5a presents an example of the nonlocal
resistance when the current flows between contacts 4-8 and the
voltage is measured between contacts 5-7, i.e. $R_{I=4,8; V=5,7}$.
The nonlocal resistance peaks are narrower and lower as compared
to the local resistance peaks measured in the same device. Simple
estimation from Kirchoff formula (see for details [16]), gives
$R_{I=4,8; V=5,7}=0.6\frac{h}{e^{2}}$ for the mean free path
$l=10\mu m$, which in 2 times large than the experimental value
$R_{I=4,8; V=5,7}=0.3\frac{h}{e^{2}}$. It is not surprising, since
when the edge current flows over a long distance (in this
particular case $L_{4,8} >> l$) there is a high probability for
the coupling with the bulk states, and the total current
experiences considerable leakage into the bulk. An advanced
theory, considered in [16], is required for a more detailed
analysis. However, we may conclude here that the edge state
transport dominates in the voltage interval $-1 V < V_{g} < 1V$.
Figure 5b shows the coefficient
 $GS^{2}$ as a function of the gate voltage near the CNP in the edge state transport regime at T=4.2K.
 It is clear that an enhancement of  $GS^{2}$ is observed compared to other than TI cases on the hole-side of the resistance peak.
 Unfortunately this enhancement is observed at low temperature and is unlikely to survive at high T ( see figure 1).
 However, quite recently the TI state has been observed at $T \sim 100K$ [22], and one may hope that the thermopower characteristics
presented here could be valid at higher temperatures in these new
materials.

\section{Theory}

Since the thermovoltage follows almost linear temperature dependence and monotonically decreases with
increasing carrier density far away from the CNP, the thermopower is likely determined by the diffusive
mechanism. The phonon drag thermopower would have different temperature dependence. In comparison to GaAs
quantum wells, where the phonon drag mechanism is essential, in HgTe quantum wells this mechanism
should be much less important because of relative smallness of deformation-potential and piezoelectric
constants in HgTe, and because of relative smallness of the density of states of 2D carriers. In the
edge state transport, the possibility of phonon drag is negligible because of topological protection
of the edge states. Therefore, we consider the diffusive mechanism in the following.

The transport in the edge channel $k$ (since the channel is helical, $k=1,2$ denote both
the direction of motion and spin orientation) along the axis $Ox$ is described by the
Boltzmann equations for the energy distribution functions of electron, $f_{k \varepsilon}(x)$:
\begin{equation}
s_k \frac{\partial f_{k \varepsilon}}{\partial x} = \gamma_{\varepsilon}
(f_{k' \varepsilon}-f_{k \varepsilon}) + g_{\varepsilon} (F_{k \varepsilon}-f_{k \varepsilon})
+\frac{J_{k \varepsilon}^{ee}}{v}+\frac{J_{k \varepsilon}^{ph}}{v},
\end{equation}
where $k' \neq k$, $v$ is the edge state velocity, $F_{k \varepsilon}(x,y)$ is the
isotropic part of the electron distribution function in the bulk,
$\gamma_{\varepsilon}=\nu^{bs}_{\varepsilon}/v$ and $g_{\varepsilon}=\nu^{eb}_{\varepsilon}/v$
are the inverse mean free path lengths for elastic backscattering and edge to bulk scattering
($\nu^{bs}$ and $\nu^{eb}$ are the corresponding scattering rates). Next, $J^{ee}$ and $J^{ph}$ are
the collision integrals for inelastic processes, electron-electron and electron-phonon scattering.
Assuming that the electrons in the state $1$ move from the left to the right (i.e., have positive
velocity), we put $s_1=1$ and $s_2=-1$. The transitions between counterpropagating edge states
(either elastic or inelastic [23]) are rare because of topological protection, while the scattering
between the edge and the bulk is weak
because of low probability of finding the bulk-state puddles [21] in the close vicinity to the edge
(it may become strong, however, above the threshold of band conductivity). On the other hand, the
inelastic electron-electron scattering within a single edge channel is free from these restrictions
and, therefore, is much stronger. Moreover, such kind of scattering has a very large phase space,
especially if $v$ is energy-independent so that momentum and energy conservation rules are
satisfied simultaneously for any two electrons participating in the collisions [24].
As the electron-electron scattering controls electron distribution, we search the distribution function
in the Fermi-like form, $f_{k \varepsilon}(x)=\{ \exp[(\varepsilon - \varphi_k(x))/T_{ke}(x)] +1
\}^{-1}$ [24], characterized by the coordinate-dependent elecrochemical potential $\varphi_k$
and temperature $T_{ke}$. Assuming that in the bulk the different spin states are weakly
coupled, we use the similar form for $F_{k \varepsilon}(x,y)$. Indeed, in the original
Bernevig-Hughes-Zhang Hamiltonian [25] describing HgTe quantum wells the spin states are uncoupled,
the coupling appears when the spin-orbit terms are introduced [2]. The coupling in symmetric wells
is described by the bulk inversion asymmetry parameter $\Delta$ [2], and the probability ratio
of spin-flip to spin-conserving transitions is estimated as a ratio of the energy $\Delta$ to the
gap energy. This ratio is small (about $0.05$) for our quantum wells.

Integrating Eq. (2) over $\varepsilon$, then multiplying Eq. (2) by $\varepsilon$ and integrating
again, we obtain four balance equations expressing conservation of particles and energy (for details see Supplementary information).
They are solved together with the balance equations for the bulk states and boundary conditions
expressing conservation of currents and energy fluxes at the edge. We consider the case when particle
transfer between edge and bulk occurs only in the narrow regions near the contacts, while in the
most part of the edge a dynamical equilibrium is reached, when only spin currents between edge and
bulk are flowing. This corresponds to the condition $g \gg 1/L$ ( $g$ is energy averaged quantity of $g_{\varepsilon}$,
defined below) and leads to a homogeneous
distributions of temperatures and potentials at the edge:
the spin-averaged potentials and temperatures linearly depend on $x$, while the quantities describing
spin polarization, $\delta \varphi=\varphi_1-\varphi_2$ and $\delta T_e= T_{1e}- T_{2e}$, are
constants. Only this condition is relevant, because under the opposite condition $g \leq 1/L$ the
particle transfer between edge and bulk cannot have a sizeable influence on the transport.
The current carried by a single edge is given as $I_e=(e/h) \int d \varepsilon (f_{1\varepsilon}-
f_{2\varepsilon}) =(e/h) \delta \varphi$, and the thermal current (energy flux minus
$\mu I_e/e$) along the edge is $W_e=(1/h) \int d \varepsilon (\varepsilon-\mu) (f_{1\varepsilon}-
f_{2\varepsilon})= (\pi^2/3 h) T \delta T_e$, where $\mu$ and $T$ are the equilibrium chemical
potential and temperature. The total current $I=I_{bulk}+2 I_e$ and the total thermal current
$W=W_{bulk}+2 W_e$ are connected to the voltage $\Delta V$ and temperature difference $\Delta T$
between the contacts by a linear relation
\begin{eqnarray}
\left( \begin{array}{c} I \\ (e/T) W \end{array} \right)
=\frac{2e^2}{h} \hat{{\cal G}}
\left( \begin{array}{c} \Delta V \\ \Delta T/e \end{array} \right),
\end{eqnarray}
where we introduced a thermoelectric response matrix
\begin{eqnarray}
\hat{{\cal G}}=\frac{w \tilde{\sigma}}{L} \hat{M} +\hat{c} \left[\hat{c}+ L \hat{\gamma} +
\frac{L}{2} \left( \frac{w}{2 \tilde{\sigma}} \hat{M}^{-1}+ \hat{g}^{-1}  \right)^{-1}
\right]^{-1} \!\!\!\! \hat{c}.
\end{eqnarray}
The first and the second terms of this matrix describe the bulk and the edge contributions, respectively,
and the term in the round braces describes the influence of the edge to bulk leakage on the edge transport.
Here and below, $\tilde{\sigma}= \sigma/(e^2/h)$, $\sigma$ is the bulk conductivity per spin,
$L$ is the distance between the contacts, $w$ is the sample width, and the matrices are defined as
\begin{eqnarray}
\hat{\gamma}=\left( \begin{array}{cc} \gamma & \gamma_I \\ \gamma_I & \gamma_{II}
\end{array} \right),
\hat{g}=\left( \begin{array}{cc} g & g_I \\ g_I & g_{II} \end{array} \right),
\hat{c}=\left( \begin{array}{cc} 1 & 0 \\ 0 & \pi^2/3 \end{array} \right),
\end{eqnarray}
\begin{eqnarray}
\hat{M}= \frac{1}{\sigma} \left( \begin{array}{cc} \sigma & \sigma_I \\ \sigma_I  & \sigma_{II} \end{array}
\right) = \left( \begin{array}{cc} 1 & eS_b \\ eS_b & e^2 (\kappa_b/\sigma T +S_b^2) \end{array} \right).
\end{eqnarray}
The matrix $\hat{M}$ describes thermoelectric response in the bulk (for details see Supplementary information) and is expressed through the bulk
thermopower $S=S_b$ and electronic thermal conductivity $\kappa_b$. The quantities $X$, $X_{I}$, and $X_{II}$
are introduced as the averages $X \equiv \langle X_{\varepsilon}
\rangle=\int d \varepsilon (-\partial f^{(0)}_{\varepsilon}/\partial \varepsilon ) X_{\varepsilon}$,
$X_I \equiv \langle X_{\varepsilon} (\varepsilon - \mu)/T \rangle$, and
$X_{II} \equiv \langle X_{\varepsilon} (\varepsilon - \mu)^2/T^2 \rangle$, where
$f^{(0)}_{\varepsilon}=\{ \exp[(\varepsilon-\mu)/T]+1 \}^{-1}$ is the equilibrium
distribution. Equation (3) is written under an additional assumption that spin relaxation
length in the bulk exceeds the sample halfwidth $w/2$. We also neglected energy transfer
between the edge states and the lattice, which is justified, according to our estimates, in
the samples of submillimeter length. The total thermopower $S_{tot}=-(\Delta V/\Delta T)_{I=0}$
is determined by the ratio of non-diagonal to diagonal elements of  the matrix (3),
$S_{tot}=e^{-1} {\cal G}_{12}/{\cal G}_{11}$. The total conductance is $G_{tot}=I/\Delta V=(2e^2/h)
{\cal G}_{11}$.

In the general case, Eq. (3) tells us that under conditions $\tilde{\sigma} < 1$,
where the edge conductivity can be significant, the contribution due to $\hat{g}$ is not important
(unless $g w/2 < \tilde{\sigma} < 1$ which necessarily implies $w \ll L$ because $g \gg 1/L$ is assumed).
The whole contribution of the term describing edge to bulk leakage in $\hat{\cal G}$ cannot exceed
$(L/w) \tilde{\sigma} \hat{M}$ and can be neglected in the case of $\gamma L \gg 1$.
This is a manifestation of the bottleneck effect described in the Introduction. On the other hand,
when $\tilde{\sigma} \gg 1$, there is no need to take the edge transport into account.

In the case when Sommerfeld expansion for each of the energy-dependent parameters is valid,
i.e., $\hat{X} \simeq \hat{c} X + (\pi^2/3)T X' \hat{\sigma}_x$, where $X'=d X_{\mu}/d \mu$ and
$\hat{\sigma}_x=\left(\begin{array}{cc} 0 & 1 \\ 1 & 0 \end{array} \right)$, the Mott relation
$S=(\pi^2/3 e)T (G'/G)$ is valid as well. Then the second term in Eq. (3) is simplified and
leads to the edge state contribution to the conductance and thermopower as follows:
\begin{eqnarray}
G_{e}=\frac{e^{2}}{h}{\cal F}^{-1},~
{\cal F}=1 + \gamma L + \frac{(\frac{L}{w})\tilde{\sigma}gL}{(\frac{2L}{w})\tilde{\sigma}+gL},
\end{eqnarray}
\begin{eqnarray}
S_{e}=  \frac{\pi^{2}T}{3|e|}\frac{{\cal F}'}{\cal F},
\end{eqnarray}
while the total conductance and thermopower are
\begin{eqnarray}
G_{tot}=2(G_e + G_b),~S_{tot}=\frac{S_e G_e + S_b G_b}{G_e+G_b},
\end{eqnarray}
where $G_b=\sigma w/L= (e^2/h) (w/L)\tilde{\sigma}$ is the bulk conductance per spin.
Though the theory leading to Eqs. (7) and (8) initially assumes $gL \gg 1$, these equations remain
formally valid in the opposite limit $gL \ll 1$, when the edge state contribution to the conductance
is not influenced by the bulk-edge currents and $G_{e}=\frac{e^{2}}{h}(1+\gamma L)^{-1}$.
This means that Eqs. (7) and (8) can be used as a reasonable approximation for arbitrary $gL$.

The main results of the our theoretical model is given by equations 7 and 8. One can see that the function ${\cal F}$ contains two terms: the first depends on the scattering between edges and the second describing the edge to bulk leakage. Two cases can be considered:

a)	Fermi level is inside of the gap.

In this case if the bulk transport is suppressed by localization the conductance and thermopower are governed by the backscattering between the edge states.

b)	The Fermi level enters conductance (valence ) band.

The edge state conductance and thermopower are determined by equations 7 and 8.    If $ (L/w) \tilde{\sigma } \ll gL$, the presence of the edge to bulk scattering does not strongly affect either the resistance or the thermopower, because:
\begin{eqnarray}
{\cal F} = 1+\gamma L + \frac{L}{w} \tilde{\sigma}
\end{eqnarray}

This also means that even when energy dependence of $g$ is strong, the thermopower sign alteration considered in model [18] does not occur.
For the long narrow sample and  for high mobility bulk carriers (high conductivity),
if $(L/w)\tilde{\sigma }\gg gL$ the edge state contribution to  function ${\cal F}$ should be proportional to $gL$, and the conductance and thermopower should be given by :
\begin{eqnarray}
G=\frac{e^{2}}{h}(1+\gamma L +g)^{-1}
\end{eqnarray}

\begin{eqnarray}
S_{e}=\left( \frac{\pi^{2} T}{3e}\right )\frac{{\gamma}'+{g}'}{1+\gamma L+gL}
\end{eqnarray}

Note that for the limiting  case $ {\gamma}' \ll {g}' \ll 0$ the thermopower is large and has a positive anomalous sign for electrons ( for conventional 2D and 3D system Seebek coefficient always negative for electrons). This mechanism has been predicted in model [18]. Thus, for observations of the anomalous thermopower long samples with suppressed edge to edge scattering is required.

\section{Discussion and comparison with experiment}

In the ballistic case, the edge state contribution to the
thermopower is absent. However, in the experiment we observe
thermopower signal in a quasi-ballistic case (Figs. 2 and 4).
Generally, it is expected that $\gamma$ is a smooth function of
chemical potential, monotonically decreasing away from the CNP,
because the CNP roughly corresponds to the crossing point in the
edge-state spectrum, where the transferred momentum is zero, and
an elastic scattering rate usually decreases with increasing
transferred momentum. This property of the scattering should cause
a decrease of the resistance with gate voltage $|V_g-V_{CNP}|$ and
a negative slope of the thermopower near the CNP, as it is
observed in experiment. If the energy is counted from the Dirac point, the
transferred momentum is $q = 2\varepsilon/v_{e}$. Assuming, for
example, that the backscattering rate $\nu_{\varepsilon}$ is
proportional to a Lorentzian function
$\gamma_{0}/[1+(\varepsilon/\varepsilon_{0})^{2}]$, one can find
${\gamma}'/\gamma=-(2\mu/\varepsilon_{0}^{2})/[1+(\varepsilon/\varepsilon_{0})^{2}]$.
The thermopower is a linear function of the Fermi energy near the
CNP, negative in the electron part and positive in the hole part.
If we assume that the transport near the CNP is dominated by the
edge states, then the slope of the thermopower near the CNP is
entirely determined by the energy dependence of the backscattering
rate. In this region the assumed approximation $g=0$ is relevant.
\begin{figure}[ht!]
\includegraphics[width=12cm,clip=]{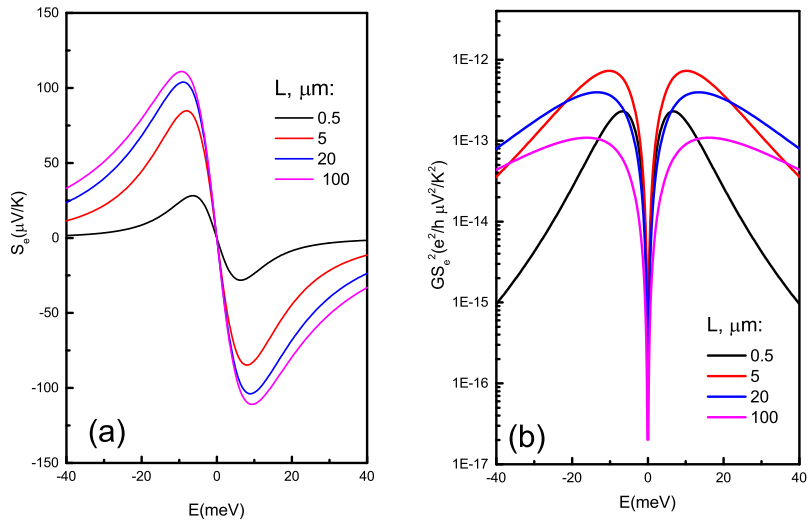}
\caption{\label{fig.6}(Color online) (a) Edge states thermopower
calculated from eqs.7 and 8 as a function of the Fermi energy for
different distances between the probes, T=4.2K. (b) Calculated
coefficient $GS^{2}$  for different distances between the probes
as a function of the Fermi energy.}
\end{figure}

Figure 6a shows the thermopower calculated from eqs.7 and 8 taking
into account the Lorentzian energy dependence of the
backscattering rate for different edge channel lengths $L$ with
the following parameters : $\varepsilon_{0}=10 meV$,
$\gamma_{0}=1/\mu m$, $\tilde{\sigma}=0.1$. One can see a
nonlinear length dependence. A nonmonotonic length dependence has
also been observed in our samples (see figure 4c). Figure 6b shows
the coefficient
 $GS^{2}$ as a function of the energy near the CNP in the edge state transport regime at T=4.2K for different edge channel lengths.
 The theory (eq.7 and 8) confirms the predictions of the model [18] concerning the dependence on geometry. In practice, however, geometry-related optimization
 would necessitate the possibility to minimize the thermoconductance of a realistic 2D TI.
 This is not feasible in our case, since the thermoconductance in our structures is mostly determined by the GaAs
substrate.
 One can see that on approaching the CNP $GS^{2}$ increases, goes through a maximum, and then reaches its minimum at the CNP, which has also been observed
in our experiment ( figure 5b). Note, however,
 that the theory is simplified and does not consider the asymmetry between electron and hole side
 near the Dirac point.  The model reproduces the key feature of the
thermopower, for example, the ambipolar behaviour of the signal,
the linear temperature dependence and the nonmonotonic dependence
on the length.

As we mentioned above, for narrow and long sample, when if $(L/w)\tilde{\sigma }\gg gL$, the contribution of the ${\gamma}'$ and ${g}'$ are equally important.
Note , however, that observation of the nonlocal resistance ( figure 5a) clearly demonstrates that the edge state transport
 inside of the mobility gap ($-1.5 V < Vg < 1.5 V$). Let us assume that $gL\sim \gamma L \sim 1$. In this case the mixing between
 edge states and the bulk becomes important for probes 3-2 $(L/w=10)$  if $\sigma \gg 0.1e^{2} /h$.  From our model (ref.16)
 we estimated approximately equal value of the bulk conductivity $\sim 0.08e^{2} /)$. Note, however, that alternatively,
 this bulk conductivity may lead to bulk mechanism of the thermopower, considered below.

We can suggest that the bulk transport is
important in our samples even in the vicinity of the CNP, despite
the existence of a considerable nonlocal resistance signifying the
presence of the edge transport. One would assume that our samples
are characterized by a significant disorder level which results in
the persistence of the bulk electron transport even inside the
gap, so that the bulk component of the conductance exists in the
whole gate voltage range. Near the center of the gap, the bulk
transport may be of a hopping variety. A signature of such
transport would be an exponential temperature dependence of the
kind $R \propto \exp[(T_{c}/T)^{1/3}]$ corresponding to the Mott
law for 2D carriers. However, we do not see such a dependence
because the bulk conductivity is short-circuited by the edge
states. On the other hand, the temperature behavior of the bulk
thermopower is not expected to change considerably at the
transition from the band transport to the hopping transport. For
the band transport $S_b \propto T$. For the hopping transport, the
thermopower was calculated in Ref. 27. Applied to 2D electrons the
result of Ref. 27 can be written as
\begin{eqnarray}
S_{b}=-\frac{\lambda}{|e|}\left [\frac{\pi-2}{\pi}(T_{c}^{2} T)^{1/3}+\frac{2\pi}{3}T\right]\left(\frac{d \ln
\rho_{\mu}}{d \mu}\right)
\end{eqnarray}
where $\lambda$ is a numerical constant of the order of 1,
$\rho_{\varepsilon}$ is the density of states, $T_{c}$ is a
characteristic temperature proportional to $1/(\rho_{\mu}
a_{0}^{2})$, and $a_{0}$ is a localization length of the electron
wavefunction. Since the Mott law is only valid for $T_{c}$
considerably larger than $T$, the first term in Eq. (10) is
important together with the second one. However, in the regime
close to the onset of the band transport the localization length
becomes large and $T_{c}$ is of the order of $T$ or smaller than
$T$, so only the second term in Eq. (10) remains important and
$S_b$ is linear in $T$.

\begin{figure}[ht!]
\includegraphics[width=12cm,clip=]{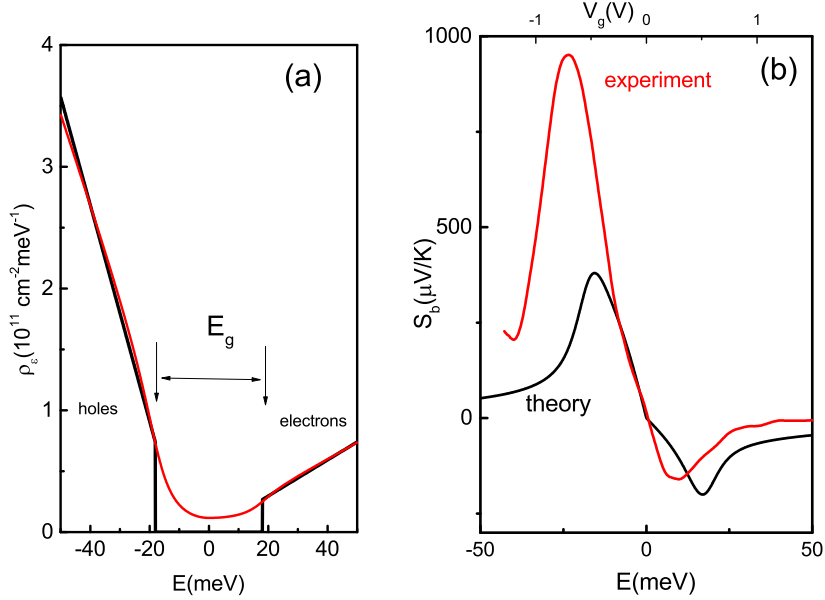}
\caption{\label{fig.7}(Color online) (a) The calculated density of
states: black line-without disorder, red line- with disorder. (b)
The calculated (hopping transport, diffusive mechanism) and
experimental thermopower at 4.2 K.}
\end{figure}

The presence of the logarithmic derivative of the density of
states is quite a general feature, because the conductivity in the
hopping transport regime is proportional to the square of the
density of states. Any physically reasonable model of the density
of states gives a dependence of S qualitatively similar to the
experimental one. We approximate the density of states by the
function:
\begin{eqnarray}
\rho_{e}=\rho_{0}+\frac{\varepsilon}{\pi A^{2}}, \,\,\,\varepsilon > 0
\end{eqnarray}
\begin{eqnarray}
\rho_{h}=\rho_{0}+\beta\frac{\varepsilon}{\pi A^{2}},\,\,\, \varepsilon < 0
\end{eqnarray}
where the energy $\varepsilon$ is counted from the center of the
gap $\varepsilon_{g}$  and $\rho_{0}$ is the constant background
density of states describing localized states. The linear function
$\varepsilon/\pi A^{2}$, where A =0.36 eV nm for a well of width d
=8 nm, describes the band density of states at $\varepsilon >
\varepsilon_{g}/2$ in the Dirac model with the electron spectrum
$\varepsilon=\sqrt{(\varepsilon_{g}/2)^{2}+(Ak)^2}$. To avoid
discontinuities at $\varepsilon = \varepsilon_{g}/2$, the function
$\varepsilon/\pi A^{2}$ is extended to the region $0 < \varepsilon
< \varepsilon_{g}/2$. The eq.(12) gives a reasonably good
description of the density of states in the electron part of the
spectrum. In the hole part, where the dispersion relation is more
complicated and may include several subbands, no simple model
exists. However, it can be roughly approximated by the same form,
with an extra coefficient $\beta$ ($\beta > 1$) describing the
electron-hole asymmetry, which leads to Eq. (13).

Any physically reasonable model of the density of states results
in a dependence of $S$ qualitatively similar to the experimental
one. For example, applying a Lorentz spectral function to describe
the broadening of electron states, one obtains

\begin{equation}
\rho_{e}=\frac{1}{\pi A^{2}}\left [\varepsilon \arctan\frac{\varepsilon-\Delta/2}{\delta}  \\
+ \varepsilon \arctan\frac{\varepsilon+\Delta/2}{\delta}+\frac{1}{2}\delta C_{\varepsilon}\right],
\end{equation}

where
\begin{eqnarray}
C_{\varepsilon}=\ln\left (\frac{(\varepsilon-E_{c})^{2}+\delta^{2}}{(\varepsilon-\Delta/)^{2}+\delta^{2}}\right)+
 \ln\left (\frac{(\varepsilon+E_{c})^{2}+\delta^{2}}{(\varepsilon+\Delta/2)^{2}+\delta^{2}}\right),
\end{eqnarray}

where $\delta$ is the broadening energy, $E_{c}$ is a large cutoff
energy. In the hole region ($\varepsilon < 0$) the extra
coefficient $\beta \approx 6 $ is added, and  the density of band
states is larger. An example of the density of states calculated
according to this expression is shown in Fig.7a with disorder
parameters $E_{c}=150$ meV and $\delta=6 meV$. Applying Eq. (11)
with a scaling constant $\lambda \approx \pi$, one may plot the
thermopower in the hopping transport regime as shown in Fig. 7b.
This calculation is in a reasonable agreement with experiment in
the region close to the CNP. The numerical constant is closer to
the metallic case. Strictly speaking, neither the hopping nor
metallic transport are expected to work well in this intermediate
regime. Still, it is important to get the idea of how the
thermopower can depend on the electron density around the CNP.
Note that the better agreement of the  experimental results with
equation (11) requires exact knowledge of the bulk conductivity
behavior in the gap and hole side regions.

The temperature dependence of thermopower does not allow to
separate the bulk transport from the edge transport, because $S_e
\propto T$ as well. Nevertheless, there are several reasons to
argue that the observed thermopower is mostly of the bulk origin.
First, the asymmetry of the thermopower signal is much larger than
the asymmetry of the resistance peak. The large asymmetry is
associated with the bulk transport and is the consequence of
electron-hole asymmetry in HgTe quantum wells (the asymmetry
increases with the width $d$ in the 2D TI regime $d > 6.3$ nm).
Second, the absolute value of $S_e$, according to Eq. (10), is
determined by the energy derivative of $\gamma$, which is not
expected to be large ($\gamma'$ is small if the backscattering
rate is weakly sensitive to the momentum transfer). In contrast,
the absolute value of $S_b$ is not small even in the region of
hopping transport, where $S_b$ is determined by the energy
derivative of the density of states according to Eqs. (14-15). In
the transition region between the hopping and band transport,
$S_b$ is expected to be large because both the density of states
and the bulk conductivity $\sigma$ strongly depend on the energy
in this region. Thus, we expect $S_e \ll S_b$. The total
thermopower in these conditions is reduced to $S \simeq
S_b/(1+G_e/G_b)$.

To conclude this chapter we should note that until now only few 2D topological insulators has been discovered:
HgTe,  InAs/GaSb based quantum wells,  and  recently – $WTe_{2}$ monolayer.  Unfortunately in all these systems $e^{2}/2h$ in
the conductance quantization has been observed in micron size devices, which indicates the presence of the backscattering between the counter propagating edge states.
 Thus, for observation of anomalous sign of the Seebek coefficient, predicted in the paper [18] it is important to develop a technology  that reduces the impurity concentration in the bulk.

\section{Summary and conclusions}

In summary, we have studied the thermoelectric power together with
the resistance behavior in HgTe quantum wells. The dependence of
the thermopower on the gate-controlled carrier density,
temperature, and device length has been investigated. The
thermopower shows a behavior expected for electron-hole systems,
changing sign at the CNP, where the resistance reaches the
maximum, and decreasing with carrier density increasing. The hole
thermopower is much stronger than the electron one. The
temperature dependence of the thermopower is close to linear.

Near the CNP, when the Fermi level lies in the gap, the resistance
is comparable to or larger than $h/2e^{2}$. The bulk of the sample
is likely to be localized under these conditions, and the
resistance is determined mostly by the edge transport. Away from
the CNP, the bulk diffusive transport takes place. In contrast,
the thermopower appears to be mostly of the bulk origin,
regardless to the position of the Fermi level. The transition from
the localized states in the gap to the band conductance does not
show the anomalies such as strong enhancement and sign variation
of the thermopower recently suggested in the theoretical works
[17,18]. Our theory of a linear thermoelectric response in 2D TI
explains the absence
of these anomalies and supports the conclusion about the bulk origin of the observed thermopower.\\

\section{Acknowledgment.}

The financial support of this work by the Russian Science Foundation (Grant No. 16-12-10041, MBE growth of HgTe QWs,
fabrication of the field effect transistors and carrying out of the experiment and data analysis) and FAPESP (Brazil), CNPq (Brazil) is acknowledged.

\title{Supplemental material: Thermoelectric transport in two-dimensional topological insulator state based on HgTe quantum well}

\author{G. M. Gusev,$^1$  O. E. Raichev,$^2$  E. B. Olshanetsky,$^3$  A. D. Levin,$^1$ Z. D. Kvon,$^{3,4}$ N. N. Mikhailov,$^3$ and S. A. Dvoretsky,$^{3}$}

\address{$^1$Instituto de F\'{\i}sica da Universidade de S\~ao
Paulo, 135960-170, S\~ao Paulo, SP, Brazil}
\address{$^2$Institute of Semiconductor Physics, NAS of
Ukraine, Prospekt Nauki 41, 03028 Kyiv, Ukraine}
\address{$^3$Institute of Semiconductor Physics, Novosibirsk
630090, Russia}
\address{$^4$Novosibirsk State University,
Novosibirsk, 630090, Russia}

\ead{gusev@if.usp.br}

\begin{abstract}
In this supplementary, we provide the
linear theory of resistance and thermopower in HgTe quantum wells in topological insulator regime.
\end{abstract}

\section{Linear theory of resistance and thermopower in HgTe quantum wells in topological insulator regime}

\subsection{Glossary}

Let us introduce a bit of terminology.

$k$ = 1, 2 - spin numbers.

$f_{k}$ - distribution function of the edge states.

$F_{k}$ - isotropic part of the distribution function of the bulk states.

These distribution functions are assumed locally Fermi-like ones, because electron-electron scattering controls the
distribution within each subsystem. For edge states this statement is proved rigorously in the Fermi liquid theory,
in helical states even better (the Luttinger liquid effects are ignored). The coordinate dependence of $f_{k}$ and $F_{k}$
enters through the parametric variables, electrochemical potentials and temperatures listed below:

$\varphi = \left( \varphi_{1} + \varphi_{2}\right)/2, \delta\varphi=\varphi_{1} - \varphi_{2}$

(below these definitions of spin-average and spin-polarized quantities apply to all variables).
$\Phi_{k}(x, y)$ - electrochemical potentials of the bulk states.

 $\Phi = \left( \Phi _ { 1 } + \Phi _ { 2 } \right) / 2 , \delta \Phi = \Phi _ { 1 } - \Phi _ { 2 }$

$\Phi_{kB}(x)$  - electrochemical potentials of the bulk states near the edge.

$\Phi _ { B } = \left( \Phi _ { 1 B } + \Phi _ { 2 B } \right) / 2 , \delta \Phi _ { B } = \Phi _ { 1 B } - \Phi _ { 2 B }$

$T_{ke}(x)$ - effective temperatures of the edge states.

$T _ { e } = \left( T _ { 1 e } + T _ { 2 e } \right) / 2 , \delta T _ { e } = T _ { 1 e } - T _ { 2 e }$

$T_{k}(x, y)$ - effective temperatures of the bulk states.

$T = \left( T _ { 1 } + T _ { 2 } \right) / 2 , \delta T = T _ { 1 } - T _ { 2 }$

$T_{kB}(x)$ - effective temperatures of the bulk states near the edge.

$\overline { T _ { B } } = \left( T _ { 1 B } + T _ { 2 B } \right) / 2 , \overline { \delta T _ { B } } = T _ { 1 B } - T _ { 2 B }$

Energy-dependent quantities:
$\sigma _ { \varepsilon } = e ^ { 2 } D _ { \varepsilon } \rho _ { \varepsilon }$- bulk energy-dependent conductivity (per spin); $D _ { \varepsilon }$ is the diffusion coefficient, $\rho _ { \varepsilon }$- is the density of
states (per spin). In zero B, bulk states are degenerate in spin, so $\sigma _ { \varepsilon }$, $D _ { \varepsilon }$ , $\rho _ { \varepsilon }$ and  are the same for both states. Total
bulk conductivity is 2 $\sigma _ { \varepsilon }$:

$\chi _ { \varepsilon } = \nu _ { \varepsilon } ^ { s } \rho _ { \varepsilon }$- spin relaxation rate in the bulk multiplied by the density of states.

$\gamma _ { \varepsilon } = \nu _ { \varepsilon } ^ { e - e } / v _ { e }$ - inverse length of free path for ”edge to edge” elastic scattering.

$g _ { \varepsilon } = \nu _ { E } ^ { e - b } / v _ { e }$- inverse length of free path for ”edge to bulk” elastic scattering.

$\upsilon _ { e }$ - velocity of edge states (constant).

In the linear theory, there appear the following averages for any energy-dependent quantity $X _ {E }$:

$X \equiv \left\langle X _ { \varepsilon } \right\rangle = \int d \varepsilon \left( - \frac { \partial f _ { \varepsilon } ^ { ( 0 ) } } { \partial \varepsilon } \right) X _ { \varepsilon }$

$X _ { I } \equiv \left\langle X _ { \varepsilon } ( \varepsilon - \mu ) / T _ { 0 } \right\rangle , X _ { I I } \equiv \left\langle X _ { \varepsilon } ( \varepsilon - \mu ) ^ { 2 } / T _ { 0 } ^ { 2 } \right\rangle$

where $f _ { \varepsilon } ^ { ( 0 ) } = 1 / \left[ \exp \left[ ( \varepsilon - \mu ) / T _ { 0 } \right] + 1 \right]$ is the equilibrium distribution, $\mu$ and $T_{0}$ are the chemical potential and the
temperature in equilibrium.
For example, when substituting the energy-dependent conductivity for $X _ { \varepsilon }$, one gets the averaged conductivity
$\sigma = \left\langle \sigma _ { \varepsilon } \right\rangle$. Further, $\sigma _ { I } = e \beta$ defines the thermoelectric coefficient$\beta$ and bulk thermopower $S = S_{b}=\beta / \sigma$. Also, $\sigma _ { I I }$ defines
the thermal conductivity $\kappa$ according to $\kappa = \left( \sigma _ { I I } - \sigma _ { I } ^ { 2 } / \sigma \right) T _ { 0 } / e ^ { 2 }$
Other parameters are obtained in a similar way. If $\sigma , X , \gamma$, and g are weakly changing in the energy interval
of T (this means, for example, $\sigma \gg \sigma ^ { \prime } T _ { 0 } \gg \sigma ^ { \prime \prime } T _ { 0 } ^ { 2 }$ and so on), the Sommerfeld expansions are possible:
$X _ { I } \simeq \left( \pi ^ { 2 } / 3 \right) X ^ { \prime } T _ { 0 } , X _ { I I } \simeq \left( \pi ^ { 2 } / 3 \right) X$, where $X ^ { \prime }$ denotes the energy derivative of $X _ { \varepsilon }$ at $\varepsilon = \mu$. This leads to Mott
relation $S \simeq \left( \pi ^ { 2 } / 3 e \right) \left( \sigma ^ { \prime } / \sigma \right) T _ { 0 }$ and Widemann-Frantz law $\varepsilon \simeq \sigma _ { I I } T _ { 0 } / e ^ { 2 } \simeq \left( \pi ^ { 2 } / 3 e ^ { 2 } \right) \sigma T _ { 0 } \equiv \kappa _ { 0 }$.

\subsection{Currents}
In this subsection we describe the basic equations for current, which we used in the theory.
Bulk current density per spin:
\begin{equation}
\mathbf { j } _ { k } = - e ^ { - 1 } \sigma \left( \nabla \Phi _ { k } + e S \nabla T _ { k } \right)
\end{equation}

Bulk energy density flow per spin:
\begin{equation}
\mathbf { H } _ { k } = ( \Phi + e S T ) \mathbf { j } _ { k } / e - \kappa \nabla T _ { k }
\end{equation}
Since the theory is linear, $\Phi$ and T in this expression can be replaced by $\mu$ and $T_{0}$.
Edge current (for a single edge). It is assumed that the state 1 moves in positive direction and the state 2 in
negative direction:

\begin{equation}
I _ { e } = \frac { e } { h } \int d \varepsilon \left( f _ { 1 \varepsilon } - f _ { 2 \varepsilon } \right) = \frac { e } { h } \delta \varphi
\end{equation}

Edge energy density flow:
\begin{equation}
H _ { e } = \frac { 1 } { h } \int d \varepsilon \varepsilon \left( f _ { 1 \varepsilon } - f _ { 2 \varepsilon } \right) = \frac { 1 } { h } \left[ \varphi \delta \varphi + \frac { \pi ^ { 2 } } { 3 } T _ { e } \delta T _ { e } \right]
\end{equation}
In the linear theory $\varphi$ and $T_{e}$ can be replaced by $\mu$ and $T_{0}$.
Total electric current along x in the Hall bar of width $w$:
\begin{equation}
J = \int _ { 0 } ^ { w } d y \mathbf { j } + 2 I _ { e }
\end{equation}
Total energy flow:

\begin{equation}
H = \int _ { 0 } ^ { w } d y \mathbf { H } + 2 H _ { e }
\end{equation}

where $\mathbf { j } = \mathbf { j } _ { 1 } + \mathbf { j } _ { 2 } , \mathbf { H } = \mathbf { H } _ { 1 } + \mathbf { H } _ { 2 }$. J is independent of x. H is independent of x under condition of no dissipation
of energy by phonons. However, both $I_{e}$ and $H_{e}$ can depend on coordinate x in the presence of particle exchange between the bulk and the edge.

\subsection{Equations}
In this subsection we describe the basic equations derived from kinetic equations, which we used in the theory.

1. For edge states (propagating along x):

\begin{equation}
\frac { \partial \varphi } { \partial x } = - ( \gamma + g / 2 ) \delta \varphi - \left( \gamma _ { I } + g _ { I } / 2 \right) \delta T _ { e } + ( g / 2 ) \delta \Phi _ { B } + \left( g _ { I } / 2 \right) \delta T _ { B },
\end{equation}

\begin{equation}
\frac { \partial \delta \varphi } { \partial x } = - 2 g \left( \varphi - \Phi _ { B } \right) - 2 g _ { I } \left( T _ { e } - T _ { B } \right),
\end{equation}
\begin{equation}
\frac { \pi ^ { 2 } } { 3 } \frac { \partial T _ { e } } { \partial x } = - \left( \gamma _ { I } + g _ { I } / 2 \right) \delta \varphi - \left( \gamma _ { I I } + g _ { I I } / 2 \right) \delta T _ { e } + \left( g _ { I } / 2 \right) \delta \Phi _ { B } + \left( g _ { I I } / 2 \right) \delta T _ { B },
\end{equation}

\begin{equation}
\frac { \pi ^ { 2 } } { 3 } \frac { \partial \delta T _ { e } } { \partial x } = - 2 g _ { I } \left( \varphi - \Phi _ { B } \right) - 2 g _ { I I } \left( T _ { e } - T _ { B } \right).
\end{equation}

Four 1-st order equations. Require 4 boundary confitions.
These equations are derived from kinetic equations. The first pair satisfies the current (particle flow) conservation,
the second one for energy flow conservation. In the second pair, the dissipation of edge state energy by the lattice
(through electron-phonon interaction) is neglected. Estimates was made to justify this neglect in 10$\mu$m size samples.
In centimeter size samples one needs to take the dissipation into account.
2. In the bulk:
Particle flow and energy flow conservation impose equations for bulk temperatures and electrochemical potentials.
The total currents and energy flows satisfy obvious relations
\begin{equation}
\nabla \cdot \mathbf { j } = 0 , \quad \nabla \cdot \mathbf { H } + P ^ { p h } = 0,
\end{equation}

where $P^{ph}$ is the power lost to phonons. More detailed analysis is needed for partial flows $j_{k}$ and $G_{k}$. There
must be taken into account particle exchange between states 1 and 2 due to spin-flip scattering and the energy
exchange between these states due to spin-flip scattering, electron-electron interaction, and electron-phonon interaction
[corresponding isotropic collision integrals are $J _ { k } ^ { s } ( \varepsilon ) , J _ { k } ^ { e e } ( \varepsilon )$ , and $ J _ { k } ^ { p h } ( \varepsilon )$.
 Writing the collision integrals for spin-flip as
 $J _ { 1,2 } ^ { s } ( \varepsilon ) = \mp \frac { 1 } { 2 } \nu _ { \varepsilon } ^ { s } \left( F _ { 1 \varepsilon } - F _ { 2 \varepsilon } \right)$
 and assuming $\int d \varepsilon \rho _ { \varepsilon } \varepsilon J _ { 1,2 } ^ { e e } ( \varepsilon ) = \mp R _ { e e } \left( T _ { 1 } - T _ { 2 } \right) ,$ and $P _ { k } ^ { p h } = - \int d \varepsilon \rho _ { \varepsilon } \varepsilon J _ { k } ^ { p h } ( \varepsilon ) = R _ { p h } \left( T _ { k } - T _ { L } \right)$
 ($T_{L}$ is lattice temperature), we obtain:
 \begin{equation}
\nabla ^ { 2 } \Phi + e S \nabla ^ { 2 } T = 0,
\end{equation}

  \begin{equation}
\kappa \nabla ^ { 2 } T = R _ { p h } \left( T - T _ { L } \right),
\end{equation}

  \begin{equation}
\nabla ^ { 2 } \delta \Phi + e S \nabla ^ { 2 } \delta T = \Gamma ^ { 2 } [ \delta \Phi + \eta \delta T ],
\end{equation}

\begin{equation}
S \nabla ^ { 2 } \delta \Phi + \xi \nabla ^ { 2 } \delta T = \Gamma ^ { 2 } [ \eta \delta \Phi + \zeta \delta T ].
\end{equation}

where $\xi = e ^ { 2 } \left( \kappa + \sigma S ^ { 2 } T _ { 0 } \right) / \left( \sigma T _ { 0 } \right) = \sigma _ { I I } / \sigma$ is a dimensionless quantity which is equal to
$\pi ^ { 2 } / 3$  if Widemann-Frantz law
works. Next, $\Gamma = \sqrt { e ^ { 2 } } \chi / \sigma$ is the inverse length of spin relaxation, $\eta = \chi _ { I } / \chi ,$ and $\zeta = \chi _ { I I } / \chi + \left( 2 R _ { e e } + R _ { p h } \right) / \chi I _ { 0 }$ are
dimensionless quantities related to spin relaxation and energy exchange.
Four 2-nd order equations. Require 8 boundary conditions.
\subsection{Boundary conditions}

In this subsection we describe the boundary conditions, which we used in the theory.

1. Outside the contacts. The total currents and energy flow normal to the boundary should be zero:
\begin{equation}
\mathbf { n } \cdot \left. \mathbf { j } _ { k } \right| _ { b o u n d a r y } = \frac { e } { h } \left[ g \left( \varphi _ { k } - \Phi _ { k B } \right) + g _ { I } \left( T _ { k e } - T _ { k B } \right) \right]
\end{equation}

\begin{equation}
\mathbf { n } \cdot \left. \left( \mathbf { H } _ { k } - \mu \mathbf { j } _ { k } / e \right) \right| _ { b o u n d a r y } = \frac { 1 } { h } T _ { 0 } \left[ g _ { I } \left( \varphi _ { k } - \Phi _ { k B } \right) + g _ { I I } \left( T _ { k e } - T _ { k B } \right) \right]
\end{equation}

where n is a unit vector perpendicular to the boundary and directed inside the sample. Since k takes two values,
there are 4 boundary conditions per each boundary.
Assuming the lower boundary at y = 0 (because we already assumed that the edge states are propagating along
x), we rewrite the boundary conditions as:

\begin{equation}
\left[ \frac { \partial \Phi } { \partial y } + e S \frac { \partial T } { \partial y } \right] _ { y = 0 } = - \tilde { \sigma } ^ { - 1 } \left[ g \left( \varphi - \Phi _ { B } \right) + g _ { I } \left( T _ { e } - T _ { B } \right) \right],
\end{equation}

\begin{equation}
\left[ e S \frac { \partial \Phi } { \partial y } + \xi \frac { \partial T } { \partial y } \right] _ { y = 0 } = - \tilde { \sigma } ^ { - 1 } \left[ g _ { I } \left( \varphi - \Phi _ { B } \right) + g _ { I I } \left( T _ { e } - T _ { B } \right) \right],
\end{equation}

\begin{equation}
\left[ \frac { \partial \delta \Phi } { \partial y } + e S \frac { \partial \delta T } { \partial y } \right] _ { y = 0 } = - \tilde { \sigma } ^ { - 1 } \left[ g \left( \delta \varphi - \delta \Phi _ { B } \right) + g _ { I } \left( \delta T _ { e } - \delta T _ { B } \right) \right],
\end{equation}

\begin{equation}
\left[ e S \frac { \partial \delta \Phi } { \partial y } + \xi \frac { \partial \delta T } { \partial y } \right] _ { y = 0 } = - \tilde { \sigma } ^ { - 1 } \left[ g _ { I } \left( \delta \varphi - \delta \Phi _ { B } \right) + g _ { I I } \left( \delta T _ { e } - \delta T _ { B } \right) \right],
\end{equation}
where $\tilde { \sigma } = \sigma / G _ { 0 }$  defines conductivity in the units of fundamental conductance quantum $G_{0} = e^{2}/h$.
2. At the contacts. If contact a is at $x = x_{a}, y = y_{a}$, contact b is at $x = x_{b}, y = y_{b}$ and the electrons in the state 1
move from a to b (in the state 2 move from b to a):
Edge:
\begin{equation}
\varphi _ { 1 } \left( x _ { a } \right) = e V _ { a } , \quad \varphi _ { 2 } \left( x _ { b } \right) = e V _ { b }
\end{equation}

\begin{equation}
T _ { 1 e } \left( x _ { a } \right) = T _ { a } , \quad T _ { 2 e } \left( x _ { b } \right) = T _ { b },
\end{equation}

Total 4 boundary conditions.
Bulk (full spin mixing and thermalization at each contact):
\begin{equation}
\Phi _ { 1 } \left( x _ { a } , y _ { a } \right) = \Phi _ { 2 } \left( x _ { a } , y _ { a } \right) = e V _ { a } , \quad T _ { 1 } \left( x _ { a } , y _ { a } \right) = T _ { 2 } \left( x _ { a } , y _ { a } \right) = T _ { a },
\end{equation}

\begin{equation}
\Phi _ { 1 } \left( x _ { b } , y _ { b } \right) = \Phi _ { 2 } \left( x _ { b } , y _ { b } \right) = e V _ { b } , \quad T _ { 1 } \left( x _ { b } , y _ { b } \right) = T _ { 2 } \left( x _ { b } , y _ { b } \right) = T _ { b },
\end{equation}

where $V_{a}, T_{a}, V_{b}, T_{b}$ are voltages and temperatures at the contacts.
4 boundary conditions per contact.
\subsection{Solutions}
In this subsection we show the solutions of the differential equations with boundary conditions described above.
Note, that we show the final solutions and discussion in the main text.

In general, numerical solution is required.
If edge to bulk scattering is ignored (g = 0), analytical solution is possible because edge and bulk subsystems are
not coupled. Assume two contacts at x = 0 and x = L, with voltage and temperature differences between them $\Delta V$
and $\Delta T $, and boundaries at y = 0 and y = w. We can assume a constant temperature gradient (the same for electrons
and lattice), so the bulk solution is straightforward $( \delta \Phi = 0 , \delta T = 0 , \Phi = \Phi ( 0 ) - e \Delta V x / L , T = T ( 0 ) - \Delta T x / L )$ and
the current is $J = 2 \sigma w ( \Delta V + S \Delta T ) / L$. For edge states the solution is $\delta \varphi = c o n s t , \delta T _{e}=const$, and the current is

\begin{equation}
I _ { e } = G _ { e } \left( \Delta V + S _ { e } \Delta T \right),
\end{equation}

where $G_{e}$ and $S_{e}$ are edge conductance and thermopower:
\begin{equation}
G _ { e } = \frac { 1 } { R _ { e } } , R _ { e } = \frac { h } { e ^ { 2 } } \left[ 1 + \gamma L - \frac { \left( \gamma _ { I } L \right) ^ { 2 } } { \pi ^ { 2 } / 3 + \gamma _ { I I } L } \right]
\end{equation}

\begin{equation}
S _ { e } = \frac { 1 } { | e | } \frac { \gamma _ { I } L } { 1 + \left( 3 / \pi ^ { 2 } \right) \gamma _ { I I } L }
\end{equation}

Parameter $\gamma_{\varepsilon}$ describes edge backscattering. It is very unlikely that this parameter strongly changes with energy
at the temperature scale. Therefore, one can apply the Sommerfeld expansion, which leads to:

\begin{equation}
G _ { e } = \frac { e ^ { 2 } } { h } \frac { 1 } { 1 + \gamma L }
\end{equation}

\begin{equation}
S _ { e } = \frac { \pi ^ { 2 } T  } { 3 | e | } \frac { \gamma ^ { \prime } L } { 1 + \gamma L },
\end{equation}

$S_{e}$ and $G_{e}$ satisfy the Mott relation $S _ { e } = - \left( \pi ^ { 2 } T  / 3 | e | \right) \left( G _ { e } ^ { \prime } / G _ { e } \right)$.

If edge to bulk scattering is persistent (finite g) the problem is much more complicated. For local transport (current
passes along x) it can be solved analytically in the case $gL \gg 1$, because edge to bulk particle transfer in this case
occurs only in the short regions near the contacts, while in the most part of the sample edge a dynamical equilibrium
is reached, when only spin currents between edge and bulk are flowing. This assumes a homogeneous distributions of
temperatures and potentials at the edge: the spin-average quantities linearly depend on x, $\varphi ( x ) = \Phi _ { B } ( x ) = \phi _ { 0 } - \phi x / L$,
$T _ { e } ( x ) = T _ { B } ( x ) = t _ { 0 } - t x / L$, while the spin-polarized quantities $\delta \varphi , \delta \Phi _ { B } , \delta T _ { e }$ and $\delta T _ { B }$ are constants. The constants
$\phi_{0}$ and $t_{}$ are nor essential, while $\phi = e \Delta V - \delta \varphi$ and $t = \Delta T - \delta T _ { e }$ according to the boundary conditions (22) and
(23). The behavior of $\Phi_{B}$ and $T_{B}$ is consistent with homogeneous behavior of the spin-average bulk potentials, which
follows from the first pair of bulk eqations [Eqs. (12), (13)] if local thermal equilibrium with the lattice takes place,
$T(x, y) = T_{L}(x, y)$. Thus, in the bulk we need to solve only the second pair of equations, where $\delta \Phi$ and $\delta T$ depend
only on the transverse coordinate y. Moreover, this depenence is antisymmetric with respect to y = w/2 (symmetry
axis of the sample), so it is enough to apply boundary conditions for one edge (y = 0) only. The bulk solution is
searched in the form:
\begin{equation}
\delta \Phi ( y ) = C _ { 1 } \sinh \lambda _ { 1 } ( y - w / 2 ) + C _ { 2 } \sinh \lambda _ { 2 } ( y - w / 2 ),
\end{equation}

\begin{equation}
\delta T ( y ) = C _ { 1 } b _ { 1 } \sinh \lambda _ { 1 } ( y - w / 2 ) + C _ { 2 } b _ { 2 } \sinh \lambda _ { 2 } ( y - w / 2 ),
\end{equation}

where $\lambda_{1}$ and  $\lambda_{2}$ are positive solutions of the biquadratic equation

\begin{equation}
\lambda ^ { 4 } \left( \xi - e ^ { 2 } S ^ { 2 } \right) - \lambda ^ { 2 } \Gamma ^ { 2 } ( \zeta + \xi - 2 e S \eta ) + \Gamma ^ { 4 } \left( \zeta - \eta ^ { 2 } \right) = 0,
\end{equation}

and
\begin{equation}
b _ { j } = - \frac { \lambda _ { j } ^ { 2 } - \Gamma ^ { 2 } } { e S \lambda _ { j } ^ { 2 } - \Gamma ^ { 2 } \eta }
\end{equation}

From Eqs. (31) - (34), one can find the matrix relation:
\begin{equation}
\left( \begin{array} { c } { \nabla _ { y } \delta \Phi } \\ { \nabla _ { y } \delta T } \end{array} \right) _ { y = 0 } = - \hat { \mathcal { A } } \left( \begin{array} { c } { \delta \Phi _ { B } } \\ { \delta T _ { B } } \end{array} \right)
\end{equation}

with

\begin{equation}
\hat { \mathcal { A } } = \frac { 1 } { b _ { 2 } - b _ { 1 } } \left( \begin{array} { c c } { u _ { 1 } b _ { 2 } - u _ { 2 } b _ { 1 } } & { u _ { 2 } - u _ { 1 } } \\ { b _ { 1 } b _ { 2 } \left( u _ { 1 } - u _ { 2 } \right) } & { u _ { 2 } b _ { 2 } - u _ { 2 } b _ { 1 } } \end{array} \right) ,
\end{equation}

\begin{equation}
u _ { j } = \lambda _ { j } \coth \left( \lambda _ { j } w / 2 \right)
\end{equation}

Thus, one has a set of 4 linear equations given below in the matrix form:
\begin{equation}
\left( \hat { 1 } + \hat { g } ^ { - 1 } \hat { M } \hat { \mathcal { A } } \right) \left( \begin{array} { c } { \delta \Phi _ { B } } \\ { \delta T _ { B } } \end{array} \right) = \left( \begin{array} { c } { \delta \varphi } \\ { \delta T _ { e } } \end{array} \right),
\end{equation}

and
\begin{equation}
[ \hat { c } + L \hat { \gamma } + ( L / 2 ) \hat { g } ] \left( \begin{array} { c } { \delta \varphi } \\ { \delta T _ { e } } \end{array} \right) - ( L / 2 ) \hat { g } \left( \begin{array} { c } { \delta \Phi _ { B } } \\ { \delta T _ { B } } \end{array} \right) = \left( \begin{array} { c } { e \Delta V } \\ { \left( \pi ^ { 2 } / 3 \right) \Delta T } \end{array} \right),
\end{equation}

where we defined the following 2x2 matrices:
\begin{equation}
\hat { M } = \tilde { \sigma } \left( \begin{array} { c c } { 1 } & { e S_{b} } \\ { e S_{b} } & { \xi } \end{array} \right) , \hat { g } = \left( \begin{array} { c c } { g } & { g _ { I } } \\ { g _ { I } } & { g _ { I I } } \end{array} \right) , \hat { \gamma } = \left( \begin{array} { c c } { \gamma } & { \gamma _ { I } } \\ { \gamma _ { I } } & { \gamma _ { I I } } \end{array} \right) , \hat { c } = \left( \begin{array} { c c } { 1 } & { 0 } \\ { 0 } & { \pi ^ { 2 } / 3 } \end{array} \right),
\end{equation}

and $\hat { g } ^ { - 1 }$ means inverted matrix.

A more detailed analysis is done below in the limit of narrow samples, when $\lambda _ { j } w \ll 1$. This means that neither
direct spin-flip scattering nor energy exchange between bulk states 1 and 2 can influence the distribution of potential
and temperature of bulk states. In this approximation, $\hat { \mathcal { A } } = ( 2 / w ) \hat { 1 }$. The expressions for potential and temperature
differences in the edge states is

\begin{equation}
\left( \begin{array} { l } { \delta \varphi } \\ { \delta T _ { e } } \end{array} \right) = \hat { K } ^ { - 1 } \left( \begin{array} { c } { e \Delta V } \\ { \left( \pi ^ { 2 } / 3 \right) \Delta T } \end{array} \right),
\end{equation}
 where
\begin{equation}
\hat { K } = \hat { c } + L \hat { \gamma } + \frac { L } { 2 } \left[ ( w / 2 ) \hat { M } ^ { - 1 } + \hat { g } ^ { - 1 } \right] ^ { - 1 }.
\end{equation}

The problem is reduced to finding the matrix inversion of $\hat { K }$. Then, expressing the current Eq. (3) through $\delta \varphi$ and
$\delta T_{e}$ found from Eqs. (41) and (42), one can find the conductance and thermopower for edge states. Expressing the
energy flow Eq. (4), one finds also the thermal conductivity.
The matrix inversion leads to large and complicated expressions not given here. Instead, we present the results
obtained in approximation of slow energy dependence of $\sigma$ g, and $\gamma$, when Sommerfeld expansions are valid for all
quantities, $\sigma$ g, and $\gamma$. In this case,
\begin{equation}
G _ { e } \simeq \frac { e ^ { 2 } } { h } \mathcal { F } ^ { - 1 } , \quad \mathcal { F } = 1 + \gamma L + \frac { ( L / w ) \tilde { \sigma } g L } { ( 2 L / w ) \tilde { \sigma } + g L },
\end{equation}

\begin{equation}
S _ { e } \simeq \frac { \pi ^ { 2 } T } { 3 | e | } \frac { \mathcal { F } ^ { \prime } } { \mathcal { F } }.
\end{equation}

In the main text we compare the theoretical results with experiment.

\end{document}